\documentclass[prb,preprint]{revtex4-1}
\usepackage{graphicx}

\begin{document}

\title{Strain induced magnetic domain evolution and spin re-orientation transition in epitaxial manganite films}

\author{Gyanendra Singh}
   \altaffiliation[G.S. and P.K.R. contributed equally to this work.]\\
\author{P. K. Rout}
   \altaffiliation[G.S. and P.K.R. contributed equally to this work.]\\
\author{Rajni Porwal}
\affiliation{Condensed Matter - Low Dimensional Systems Laboratory, Department of Physics, Indian Institute of Technology Kanpur, Kanpur - 208016, India}

\author{R. C. Budhani}
\email{rcb@iitk.ac.in, rcb@nplindia.org}
\affiliation{Condensed Matter - Low Dimensional Systems Laboratory, Department of Physics, Indian Institute of Technology Kanpur, Kanpur - 208016, India}
\affiliation{National Physical Laboratory, New Delhi-110012 India}

\begin{abstract}

The evolution of magnetic domain structure in epitaxial La$_{0.625}$Ca$_{0.375}$MnO$_3$ films on (001) NdGaO$_3$ is monitored as a function of temperature and magnetic field using Magnetic Force Microscopy. We see two distinct regions of magnetic orientational order; one in-plane displaying contrast-less image and the other tilted away from the film plane forming a distinct stripe pattern. A strong domain splitting is observed at the boundary of two regions, which is resilient to reorientation with temperature and magnetic field. We propose a model magnetic free energy functional to explain the mechanism of domain splitting seen in manganite films.

\end{abstract}
\maketitle

Magnetic domains (MDs) in ferromagnetic films arise from the requirement of minimization of the total magnetic free energy consisting of magnetic interactions of both local and non-local nature.\cite{Hubert} The epitaxial films of La$_{1-x}$Sr$_{x}$MnO$_3$ (LSMO) and La$_{1-x}$Ca$_{x}$MnO$_3$ (LCMO) provide suitable platforms to see rich magnetic textures because of the coupling between the charge, spin, orbital and lattice degree of freedom,\cite{Tokura} which connect to the various components of the magnetic free energy functional, and are affected significantly by epitaxial strain. The latter is a powerful tuning parameter as these oxides can be grown epitaxially on a large number of single crystal substrates which impart varying degree of strain, depending on the lattice mismatch. Due to the magnetoelastic coupling these elastic strains can induce magnetic anisotropy in the film, whose magnitude depends upon the magnetostriction constants and the amount of stress in the film. Typically the value of stress induced anisotropy lies within 10$^{4}$-10$^{5}$ Jm$^{-3}$ for thin films of various manganites.\cite{Lofland,O'Donnell,Berndt} Moreover the intrinsic magnetocrystalline anisotropy is in the order of $\sim$ 10$^{3}$ Jm$^{-3}$, which is quite small as compared to the stress induced  anisotropy.\cite{Berndt} Thus the strain plays an important role in determining the magnetic properties of the epitaxial films. The compounds LCMO and LSMO are pseudocubic perovskites with lattice parameters in the range of 0.386-0.389 nm in the unstrained form. The commonly used substrates for epitaxial growth of these oxides are LaAlO$_3$ (LAO), SrTiO$_3$ (STO) and NdGaO$_3$ (NGO). While LAO provides an isotropic in-plane compressive strain, which results in out-of-plane easy axis and maze-like domains, the films on STO have the easy axis lying in-plane due to in-plane tensile strain and thereby form planar domains.\cite{Dho} The orthorhombicity of NGO makes the compressive strain anisotropic in the film plane and thus promoting a preferential direction for MD formation. However the imaging of MD structure of the films on NGO has given contrasting results, with observation of both in-plane and out-of-plane orientation of magnetization.\cite{Dho,Desfeux,Biswas,Israel,Piano,Llyod,Vlasko} The reasons for this non-uniqueness of the magnetic texture have not yet been established.

Here we report a careful study of the evolution of MD structure in La$_{0.625}$Ca$_{0.375}$MnO$_3$ films as a function of temperature ($T$) and in-plane magnetic field ($H_{\|}$) in ultra high vacuum of $\simeq$ 10$^{-10}$ Torr using Magnetic Force Microscopy (MFM) (Scanning Probe Microscope, Model:  UHV 3500, RHK Technology). We first show the manifestations of the magnetic ordering in the magnetization $M$($T$) and resistance $R$($T$) data of a 200 nm thick LCMO film grown on (001) NGO at 800$^o$C by pulsed laser ablation of a target of La$_{0.625}$Ca$_{0.375}$MnO$_3$. The details of the thin film growth are described in several earlier publications.\cite{Padhan}  The $T_{C}$ as estimated from the $R$($T$) and $M$($T$) data is $\simeq$ 260 K [Inset of Fig. 1(a)]. Figure 1(a) shows $M$($H$) loops measured with $H_{\|}$ along [100] direction, as well as in out-of-plane field $H_{\bot}$ configuration. From the shape of the loops, it is evident that $\mathbf{M}$ has both in-plane and out-of-plane components, with in-plane remanence being $\approx$30$\%$ of the saturation value as deduced from $H_{\|}$ loop.

The MFM image of the film shows a regular stripe domain pattern with the bright and dark contrast due to a quasi-periodic orientation of the component of $\mathbf{M}$ in to and out of the film plane [Fig. 1(b)]. The period of the stripe domains ($L$) is $\simeq$ 300 nm as determined by Fast Fourier Transformation (FFT) shown in the inset of Fig. 1(d) of the MFM image. Although the homogeneous stripe domain pattern observed here has been reported previously,\cite{Dho} the striking discovery of this study is the observation of the microscopic patches of two magnetically distinct phases separated by a thin boundary marked by line B in the MFM image shown in Fig. 1(c). A remarkable feature of Fig. 1(c) is the branching of the stripes as the boundary of the CL region is approached. This splitting of the domains has been quantified by taking line scans perpendicular to the length of the stripes at various points across the boundary from line A to B and the variation of $L$ with distance is shown in Fig. 1(d). While the stripes multiply, the MFM signal intensity drops on approaching the boundary. One may suspect that the CL phase is confined to the thinner sections of the film where the dominant effect of dipolar energy will try to make $\mathbf{M}$ in-plane. To address this possibility, we have done MFM imaging after Ar$^{+}$ ion milling of the film down to 100 nm but the stripes persist even at this thickness. Clearly, the effect is not due to a variation in film thickness. The simplest explanation for these observations can be based on the spatial variation of strain in the film. While a previous report on LCMO grown on (001) NGO reveals a critical thickness of $\simeq$ 500 nm for fully relaxed bulk like state,\cite{Seiro} it is possible that at intermediate thickness of $\simeq$ 200 nm of our film both strained and relaxed regions coexist due to partial strain relaxation. Although the branching of domains is expected near the edge of the sample, all CL regions observed in the film lie well within the sample boundaries. We have taken a number of scans at different places of the film and observed only $\sim$ 5$\%$ of such CL regions.

A more informative approach to understand MDs would be to see the variation of the angle $\theta$ between $\mathbf{M}$ and film plane [Fig. 1(e)], which has been extracted as $\theta$($x$,$y$) = sin$^{-1}$[$MFM$($x$,$y$)/$MFM_{max}$], where $MFM$($x$,$y$) is the intensity of the MFM signal and $MFM_{max}$ is its maximum value corresponding to $\theta$ = $\pm$90$^0$.\cite{Note1} We see two separated regions; one with $\theta \approx$ 0$^o$ (in-plane $\mathbf{M}$) and the striped region with non zero $\theta$'s. Surprisingly, near the phase boundary we see a region with maximum canting angle $\theta_{m}$ (dark red and dark blue), which is much higher than $\theta_{m} \approx$ 65$^0$ seen in the regions away from the boundary. Moreover this $\theta_{m} \approx$ 65$^0$ is distinct in the histogram extracted from the $\theta$ image [Fig. 1(f)]. The sinusoidal oscillation of $M_z$ with a particular $\theta_m$ would give an equal number of counts to the $\theta$ values ranging from $+\theta_m$ to $-\theta_m$.\cite{Note2} This clearly means a symmetric histogram profile about 0$^o$. Thus a more common value of the $\theta_m$ implies discontinuities at $+\theta_m$ and $-\theta_m$ in histogram. We indeed observe discontinuities in 63-67$^o$ range on both branches of histogram.

In order to address the micromagnetic domain evolution at the boundary region in the vicinity of $T_{C}$, the MFM images are taken at various temperatures. At $T \leq$ 240 K, both the phases with domain branching at their boundary are observed (See Fig. 2). As the temperature increases to 254 K, the branching disappears, but the stripes remain, though with attenuated intensity. On further increasing the temperature to 261 K, the stripe domains disappear completely in the interior regions; while some magnetic contrast persists at the boundary till 264 K. It is clear that the magnetic order parameter at the boundary is much more resilient to change, suggesting some kind of a pinning mechanism in action. The possible origin of such magnetic inhomogeneity can be found in Ref. [13], where mesoscale regions with different $T_C$ are observed near the artificial grain boundaries in LSMO films grown on bicrystal STO substrates. They have attributed such effect to the spatial variation in the strain, which we believe to be the reason for non-trivial domain pattern seen in our film. A similar resilience of boundary to reorient is observed for in-plane magnetic fields up to $\approx$ 100 mT.

All these observations can be explained by a simple model in which the film lies on the $xy$-plane and $\mathbf{M}$ has both $y$ and $z$ components in addition to $M_z$ oscillating as sin($\pi$$x$/$L$) as sketched in Fig. 3(a).\cite{Yafet,Wu} The free energy density $E$ of such a domain pattern can be expressed as:
\begin{eqnarray}
 E &=& \frac{{A\pi ^2 }}{{2L^2 }}(1 - \cos \theta_m ) + \frac{1}{2}(K_1 \cos ^2 \theta_m  + K_2 \cos ^4 \theta_m ) \nonumber\\
 &&+\frac{\Omega }{2}\sin ^2 \theta_m  - \frac{{\pi \Omega a_0 }}{{4L}}\sin ^2 \theta_m  - \frac{{2\mu_0HM_S}}{\pi }I(\theta_m )
\end{eqnarray}
Here the first term is the exchange interaction energy with $A =$ 1.7$\times$10$^{-12}$ Jm$^{-1}$.\cite{Lynn} The second term is the uniaxial magnetic anisotropy energy taken up to the fourth order and the next two terms are the short and long range part of dipolar energy with $\Omega = \frac{1} {2} \mu_0 M_S^2 =$ 1.01$\times$10$^{5}$ Jm$^{-3}$. The last term is the Zeeman energy for an in-plane magnetic field $H$ applied along $y$-axis with $I(\theta_m ) = \int_0^{\pi /2} {d\varphi \sqrt {1 - \sin ^2 \theta_m \sin ^2 \varphi }  \approx 1 + \left( {\frac{\pi }{2} - 1} \right)(\cos \theta_m )^{3/2} }$.

Minimization of Eq. (1) relative to $L$ and $\theta_m$ yields an analytical expression for domain period:
\begin{eqnarray}
 L = \frac{{4\pi A}}{{\Omega a_0(1 + \cos \theta_m )}}
\end{eqnarray}
where cos $\theta_m$ satisfies the relation
\begin{eqnarray}
3\frac{{\Omega ^2 a_0^2 }}{{32A}}\cos ^2 \theta_m  + \left( {K_1  + 2K_2 \cos ^2 \theta_m  - \Omega  + \frac{{\Omega ^2 a_0 ^2 }}{{16A}}} \right)\cos \theta_m \nonumber\\
- \frac{{3\mu_0HM_S}}{\pi }\left( {\frac{\pi }{2} - 1} \right)(\cos \theta_m )^{1/2}  - \frac{{\Omega ^2 a_0^2 }}{{32A}} = 0
\end{eqnarray}

The stripe domains present in our film can be explained by the canted state of $\textbf{M}$ introduced solely by a non-zero $K_2$.\cite{Note2} The least square fit shown in Fig. 1(a) of the in-plane magnetic hysteresis data yields the anisotropy values of $K_1/\Omega \approx$ 0.95 (or $K_1 \approx$ 9.60$\times$10$^{4}$ Jm$^{-3}$) and $K_2/\Omega \approx$ 0.12 (or $K_2 \approx$ 1.21$\times$10$^{4}$ Jm$^{-3}$).\cite{Note1} In comparison the value of magnetocrystalline anisotropy is at least an order of magnitude smaller. The stress induced anisotropy for biaxial strains can be expressed as $K_s  = 3\lambda \sigma /2$, where $\lambda$ is the magnetostriction constant and $\sigma$ is the stress.\cite{O'Donnell} The stress can be estimated from the product of the Young's modulus ($Y$) and the strain ($\epsilon$). The typical values of $\epsilon \sim$ 10$^{-2}$ and $Y \sim$ 10$^{11}$-10$^{12}$ Nm$^{-2}$ result in stress values of $\sim$ 10$^{9}$-10$^{10}$ Nm$^{-2}$.\cite{Berndt} Using literature value of the magnetostriction constant $\lambda \sim$ 10$^{-4}$,\cite{Ibarra} one obtains an anisotropy of $K_s \sim$ 10$^{5}$ Jm$^{-3}$, which is quite close to our value of $K_1$. This indicates that the anisotropy present in the system is predominantly due to the elastic strains. With $L \simeq$ 300 nm as calculated before, we have $\theta_{m} \approx$ 65$^o$, which is in agreement with the value determined from $\theta$ image.

Equation (2) reveals that as $\theta_m$ increases from 0$^o$ to 90$^o$, $L$ becomes twice as large. Thus, as we move from stripe domain region to CL region where $\theta_m =$ 0$^o$, one can expect a subdivision of the stripes. A simple way to explain the mechanism responsible for the change in $\theta_m$ will be a variation of magnetic anisotropy due to local elastic strains. The minimization of total anisotropy energy along the lines of Ref. [18] under the appropriate strain conditions ($\epsilon_{xx} \approx \epsilon_{yy} \neq$ 0 and $\epsilon_{xy} =$ 0) shows a direct proportionality of $K$'s with the in-plane strain, which means higher the compressive strain higher will be the anisotropy and vice versa. All the possible magnetic domain configurations are summarized in Fig. 3(b) for $K_1$/$K_2 =$ 7.9 in our film. We can clearly see that $K_1$/$\Omega =$ 0.80 is the critical anisotropy value separating planar domains from stripe ones. For $K_1$/$\Omega <$ 0.80 (region-I), \textbf{M} is completely in-plane whereas the stripe domains with canted \textbf{M} can be observed for $K_1$/$\Omega >$ 0.80 (region-II $\&$ III). The maximum canting angle $\theta_m$ gradually increases in region-II and reaches a value of 90$^o$ in region-III. A slight decrease or increase in strain from the present value ($K_1$/$\Omega =$ 0.95) can move the system towards region-I or III respectively and thus a transition between planar and stripe domain pattern as observed here can be expected. Furthermore the contradicting domain structures reported previously in Ref. [3-9] can be due to the strain present in the films depending on the growth conditions and the film thickness. Similar to the nature of $L$, we see a minor increase in $K_1/\Omega$ followed by a gradual drop whereas the corresponding $\theta_{m}$'s decrease rapidly to 0$^o$ from a value of 65$^o$ while reaching a maximum of 85$^o$ at the middle [See Fig. 3(c)]. Although phase separated regions resulting from the local variation in strain have been proposed before,\cite{Tovstolytkin} we have given a direct and visual evidence of different $\mathbf{M}$ states with same magnetic ordering temperature in LCMO film.

In summary, MFM has been used to establish the two distinct orientations of $\mathbf{M}$ in La$_{0.625}$Ca$_{0.375}$MnO$_3$ epitaxial films. A stripe domain pattern caused by a tilted orientation of $\mathbf{M}$ with respect to the film plane shows a distinct subdivision in a manganite film. A model calculation predicts the fragmentation of the stripes at the boundary due to the local variation of strain.

The authors thank P. C. Joshi for technical help. This research has been supported by grants from DIT. G.S. and P.K.R. acknowledge the financial support from CSIR, India. R.C.B. acknowledges the J. C. Bose fellowship of DST.

\pagebreak

\begin{figure}[h]
\begin{center}
\includegraphics [width=12cm]{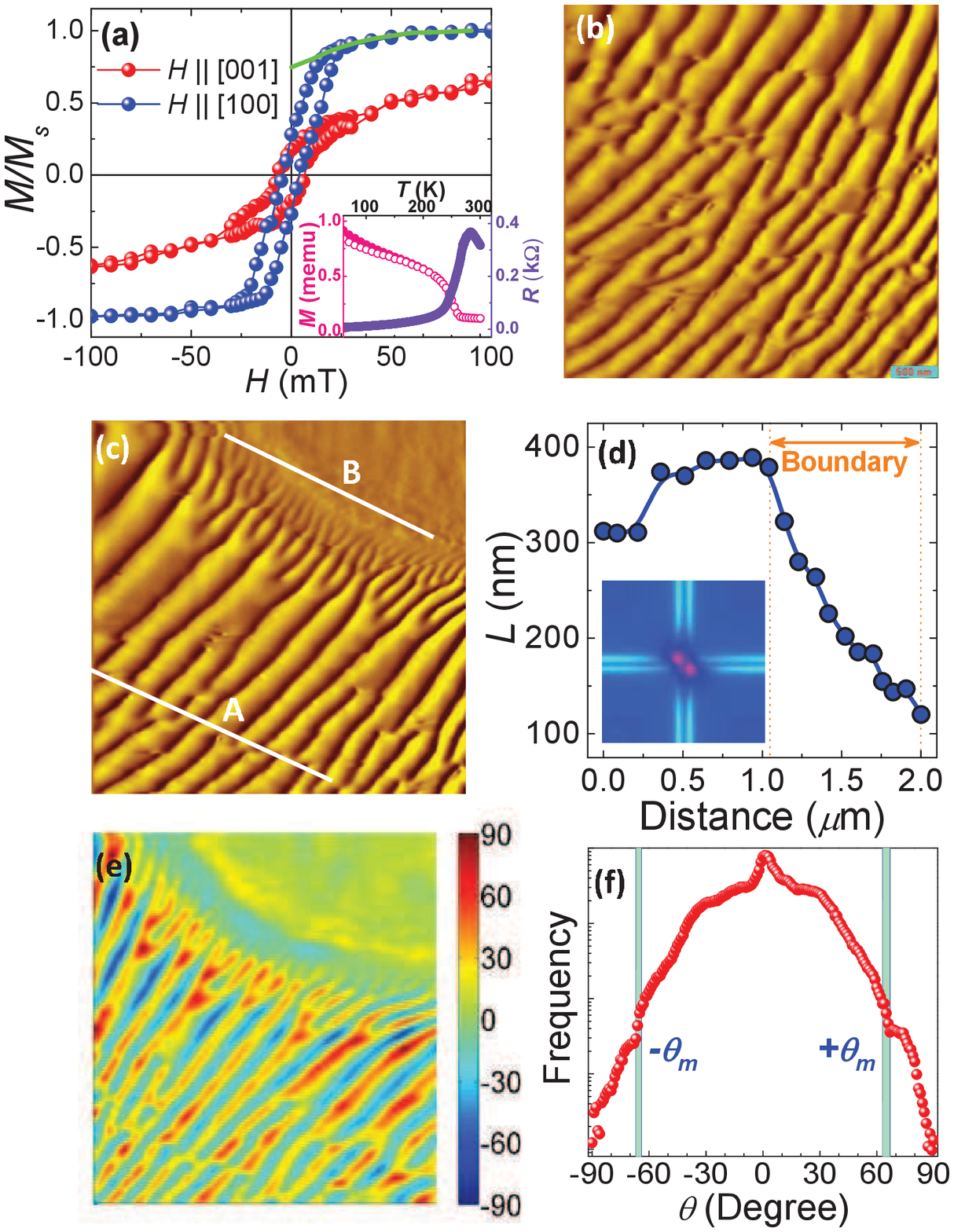}%
\end{center}
\caption{\label{fig1} (a) Hysteresis loops measured with field ($H$) along [100] and [001] directions. The solid line shows the least square fit to in-plane hysteresis loop.\cite{Note1} The inset shows the $R$ vs. $T$ along with zero field cooled (open circle) and field cooled (solid circle) $M$ vs. $T$ measured in $H =$ 20 mT. (b $\&$ c) The MFM images of LCMO film taken at two different places at 120 K in the absence of magnetic field after zero field cooling. Here the scan area is 4$\times$4 $\mu$m$^2$. (d) The domain period as a function of length measured from line A to B as shown in (c). The inset shows the FFT image of the panel (b). (e) The $\theta$ image extracted from MFM image (c). (f) The histogram of $\theta$ image showing most probable $\theta_m$ values (blue regions).}
\end{figure}
\pagebreak

\begin{figure}[h]
\begin{center}
\includegraphics [width=12cm]{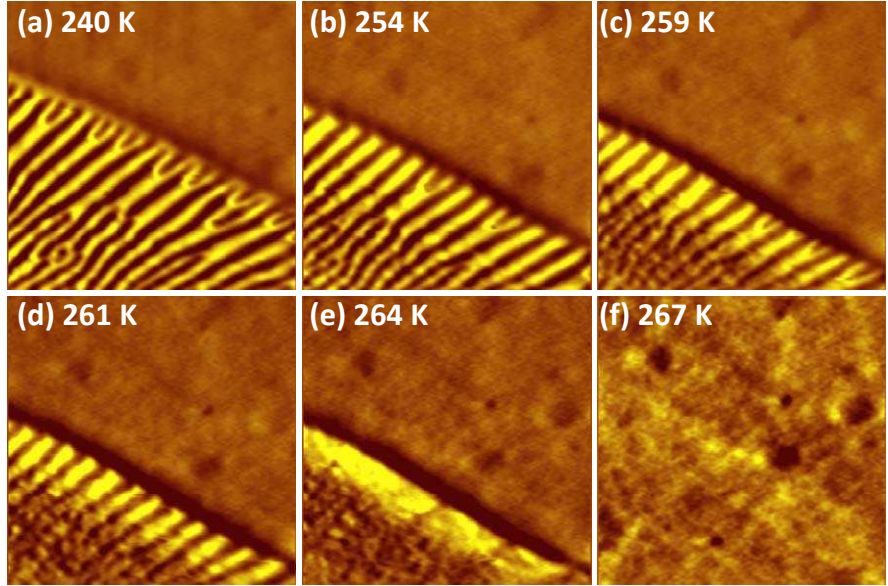}%
\end{center}
\caption{\label{fig 2} (a-f) The MFM images at several temperatures near $T_{C}$ in zero magnetic field. All images captured have the scan area of 4$\times$4 $\mu$m$^2$ taken at the same place in continuous heating mode while the nanometer scale offset due to the thermal drift has been controlled by $x$-$y$ offset.}
\end{figure}
\pagebreak

\begin{figure}[h]
\begin{center}
\includegraphics [width=12cm]{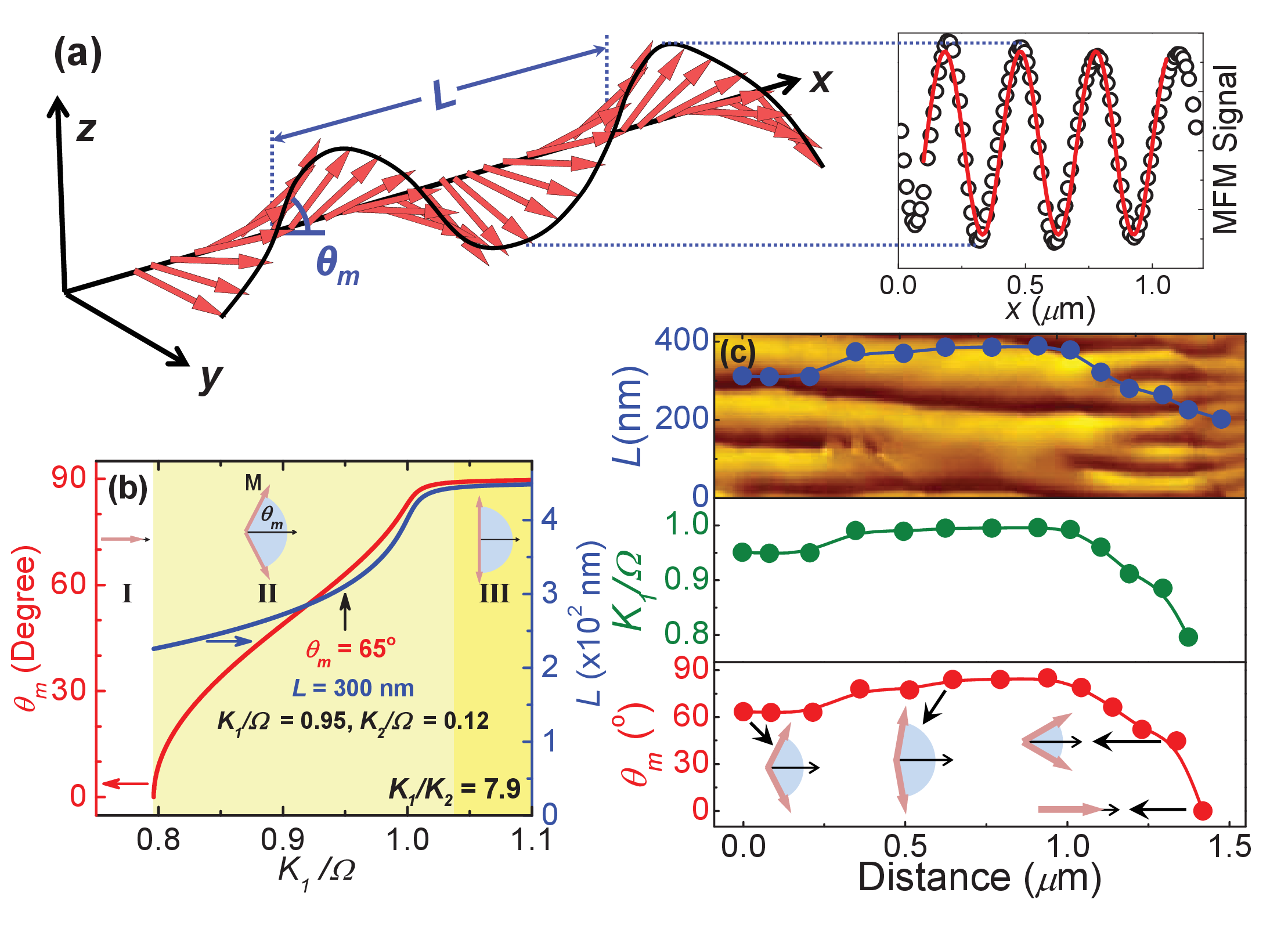}%
\end{center}
\caption{\label{fig 3} (a) Schematic image the stripe domain pattern with $\mathbf{M}$ lying completely in $yz$-plane with $M_z$($x$) $\sim$ sin($\pi x/L$). The right panel show the MFM signals as we move across the stripe length, which fits quite well with a sinusoidal function (red line). (b) The $\theta_m$ and $L$ as a function of $K_1$/$\Omega$ with $K_1$/$K_2$ = 7.9. The projections of \textbf{M} on $yz$-plane are shown for different regions. (c) The variation of $L$, $K_1$/$\Omega$ and $\theta_m$ as we move from line A to B shown in Fig. 1(c).}
\end{figure}

\end{document}